\documentclass[10pt,twocolumn]{IEEEtran}
\usepackage{multicol}

\usepackage{color}
\usepackage{soul}
\usepackage{amsmath}
\usepackage{graphics}
\usepackage{setspace}
\usepackage{cite}
\usepackage{latexsym}
\usepackage{float}
\usepackage{epsfig}
\usepackage{multirow}
\usepackage[table,xcdraw]{xcolor}
\usepackage{cite,cases,url}
\usepackage{amssymb}
\usepackage{graphicx}
\usepackage{epstopdf}
\usepackage{balance}
\usepackage{subcaption}
\usepackage{enumerate}
\usepackage{array}
\usepackage{algorithmic}
\usepackage{algorithm}
\usepackage{enumitem}
\usepackage[normalem]{ulem}
\useunder{\uline}{\ul}{}

\newcolumntype{L}{>{\centering\arraybackslash}m{5cm}}
\newcolumntype{K}{>{\centering\arraybackslash}m{6cm}}
\newcolumntype{P}{>{\centering\arraybackslash}m{2.3cm}}
\newcolumntype{M}{>{\raggedright\arraybackslash}m{2cm}}
\newcolumntype{N}{>{\raggedright\arraybackslash}m{2.5cm}}

\begin{document}

%\title{Security Aspects in 6G Radio Access Network: Threats, Solutions and Recommendations }
\title{Network and Physical Layer Attacks and Countermeasures to AI-Enabled 6G O-RAN%Radio Access Networks
}%, Countermeasures and Standardization}
\author{
\IEEEauthorblockN{Talha F. Rahman, Aly S. Abdalla, Keith Powell, Walaa AlQwider, and Vuk Marojevic}\\ \vspace{0.2cm}
\normalsize\IEEEauthorblockA{Dept. of Electrical and Computer Engineering,  Mississippi State University,
Mississippi State, MS}
%$^{\star}$Dept. Information and Communication Technologies, Universitat Pompeu Fabra, Barcelona, Spain}
}

\maketitle

\begin{abstract}
Artificial intelligence (AI) will play an increasing role in cellular network deployment, configuration and management. This paper examines the security implications of AI-driven 6G radio access networks (RANs). 
While the expected 
%e identify the expected 
timeline for 6G standardization is still several years out, %and the ongoing
pre-standardization efforts related to 6G security are already ongoing and will benefit from fundamental and experimental research. %and our recommendations. 
The Open RAN (O-RAN) describes an industry-driven open architecture and interfaces for building next generation RANs with AI control. Considering this architecture, we %provide an overview of security discuss 
identify the critical threats to data driven network and physical layer elements, the corresponding countermeasures, and the research directions. 
\end{abstract}

\IEEEpeerreviewmaketitle

\section{Introduction}
\label{sec:intro}

The %hastened enlargement of 
steady increase in the number of connected devices and the heterogeneous types of communications performance %maturity of the 
demands have driven the wireless business and %obliged for the advanced wireless communication networks and applications, serious 
research and development (R\&D) efforts. %from academia and industry for the evolution of
%on advanced wireless networks. % have been proposed. 
This led to the definition, development and evolution of cellular communications networks towards %and %Some ongoing discussions are concerning the further improvements and developments progressing beyond the 
the %currently deployed 
fifth generation (5G), which adds new degrees of flexibility and additional capacity beyond what 4G systems can provide. % since they profess it as the high-end version of wireless communication networks. 
While industry is deploying early 5G networks and standardizing new features, researchers started speculating about 6G technology and system characteristics~\cite{6GHorizons,tariq2020speculative}. %Table~\ref{tab:survey} presents the significant distinctions between the 5G and 6G communications networks %regarding 
%the %possible 
%technologies and services. %at different layers, services and applications, network architectures, and others. 
New technology %packaged into 
and redesigned networks are driven by demand. These service and quality of service demands are captured by key performance indicators (KPIs). 
%Other opinions have proved the necessities for the next upgrading of the sixth generation (6G).   
%Table~\ref{tab:KPI} compares the expectations of the important KPIs between future 5G and 6G networks. %that are commonly agreed upon by the community.

%\textcolor{red}{optional paragraph incl. Moreover:}
Compared to 5G, 6G systems are needed to satisfy the increasingly heterogeneous communications %has more strict 
requirements of applications and services, and to enable new applications and services. % and application. %through advanced wireless. 
6G networks are thus expected to offer: %Some of the envisaged requirements are %~\cite{02letaief2019roadmap}
%:
\begin{itemize}
  \item ultra-high data rates of up to 1 Tb/s;
  \item ultra-high energy efficiency;
  \item massive low-latency access and response; and
  \item extended frequency operation, from radio frequency (RF) to visible light communications. %above and below 100 GHz.
\end{itemize}
%to name but a few. 
Moreover, features such as holographic-type communications, the tactile Internet, multi-sense networks, and the digital twin, suggest that 6G networks will need to be able to support advanced IT services and technologies, notably computation oriented communications (COC), contextually agile eMBB communications (CAeC), event defined uRLLC (EDuRLLC)~\cite{6Groadmap}. 
%To be capable of fulfilling the exceptional specifications and provisions, the 6G network has to be empowered by transformative and revolutionary enablers.
% In addition to these %commonly agreed 
% requirements, we consider availability, reliability and security as the overarching concept as the key requirement for 6G to protect systems, application and data. % for two reasons: 
% Note that only a fraction of the network data will be consumed by users. %and not only this data needs to be protected, but the , the remaining will be used for system and the 
% The more complex system architecture, new hardware and increasing software dependency, % tools 6G networks will use, 
% will increase the attack surface %%make 6G networks potentially vulnerable to new attacks that
% and this needs to be contemplated as part of the design, operation, and evolution of 6G systems.

\begin{figure}[t]
    \centering
    \includegraphics[width=0.48\textwidth]{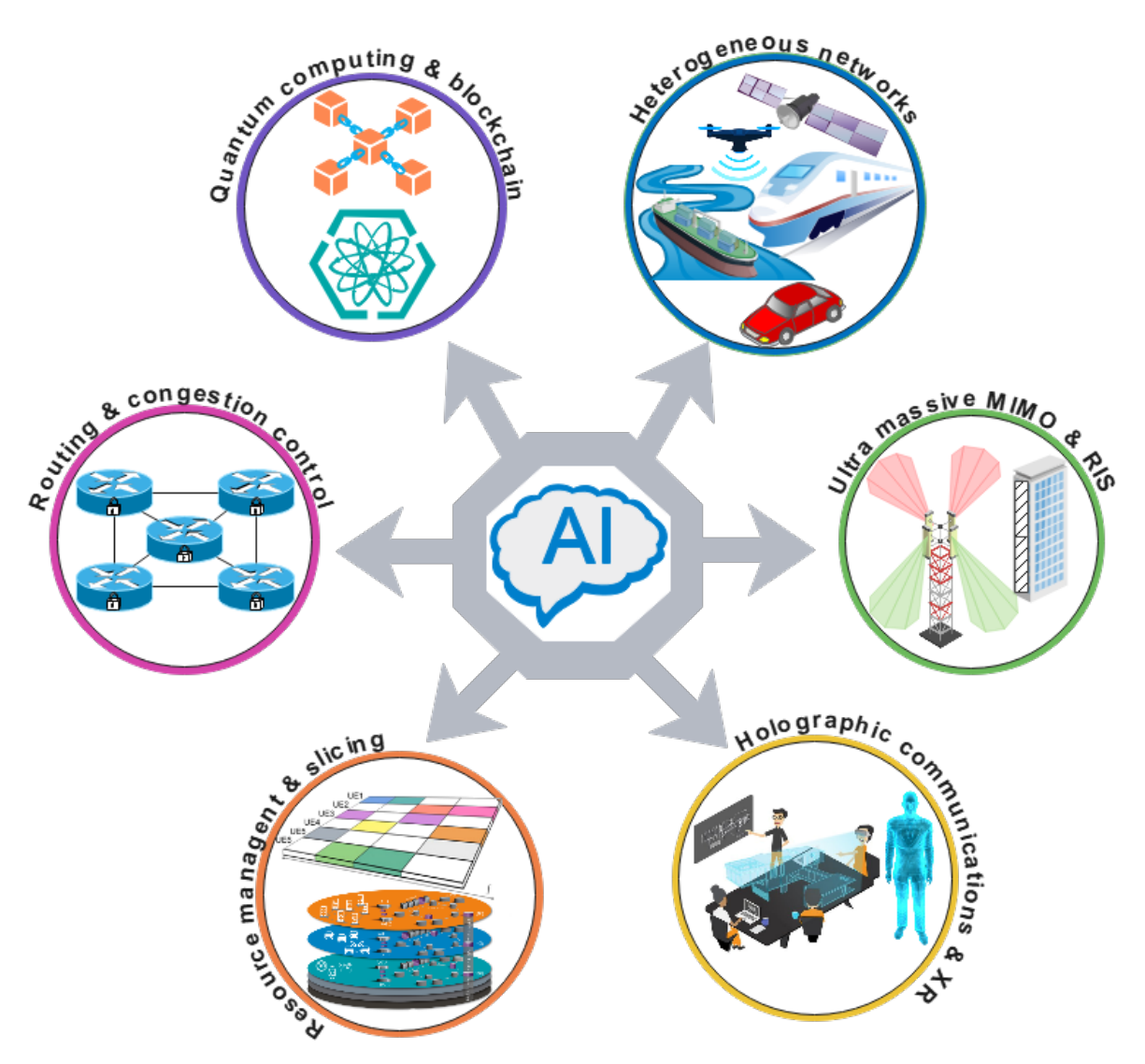}
    \caption{The features of 6G networks. (RIS: reconfigurable intellignt surface; XR: extended reality.)}
    \label{fig:Adv6G}
    %\vspace{-3mm}
\end{figure}

%As predicted in~\cite{02letaief2019roadmap}, 
%The future evolution of 
% for the aspiration of enabling complex and smart use cases. %and achieve close-to-ultimate performance. Table~\ref{tab:KPI} exhibits the divined key performance indicators (KPI's) of 6G networks compared to the achieved KIP's of 5G networks. 
Cellular networks will evolve from network softwarization %ization 
to network intelligence. 
%Additionally, the decisive component of the future 6G implementations will be the enabling intelligence via the integration of artificial intelligence and machine learning algorithms. 
The intellectual property of the 6G networks will facilitate dynamic data-driven planning and operation solutions for optimizing the real-time orchestration and management from the network edge to the core. 
6G will therefore use artificial intelligence (AI) agents to control network functions; other functions may implement AI processing algorithms.  %ed as neuronal networks as depicted in , for example.
Moreover, AI will be used to calibrate, configure and maintain the network functions as well as ensure the integrity of the subsystem, system and network. 
%Besides, the AI technologies inaugurate a chain of unorthodox and enhanced functions starting from the physical up to the application layer at both ends of the communication link for overall performance optimization. 
It can improve interference detection and mitigation, dynamic resource allocation, channel estimation and coding, scheduling, intelligent flow control, adaptive congestion control, and extensible multiple access, among others. 
%tization.
%For instance, t
%Two important features introduced in 5G---software defined networking (SDN) and network functions virtualization (NFV), have enabled evolving modern communications networks to software-based virtual networks.
Fig.~\ref{fig:Adv6G} illustrates the 6G features and technologies, empowered by AI. %will be leveraged in the 6G networks.

System integrity and security are fundamental for such systems, where hardware and software that belong to a given network, or that a network uses, will change. 
%Hence, the 6G systems that we envisage will %not
%provide the system configurations through continuous learning. 
%Moreover, such systems will challenge themselves to understand their weaknesses and make them more robust and reliable in a proactive way.
%The raised claims for 6G advancements are scrutinized based on improving the current performance and state-of-art technologies appropriated to support the 5G use cases particularly regarding the enhanced mobile broadband (eMBB), ultra-reliable and low-latency communications (uRLLC), and massive machine-type communications (mMTC), adjacent with application scenario extensions for the 6G networks  that demand spikes 
Wireless security is a fundamental feature that will drive 6G design, development, deployment and operation. %that has been the center of attention for industry and academia over all generations.
%Notwithstanding the consolidation of 5G security features corresponding to the early wireless communication generations, 
%5G networks are still prone to critical security challenges of prior generations. 
%However, 
The 6G enablers need to be leveraged for %as potentials 
enriching wireless network security. Research %enrichment 
will play a key role in minimizing the vulnerabilities of 5G and safeguarding the advanced use cases that 6G will enable. %\hl{There has been preliminary research that examined the research challenges in 6G networks such as~\mbox{\cite{6Gcontextaware,6GSecSurvey,6GSec}}. 

The works presented in~\cite{UAV,RIS} show how different 6G enablers such as unmanned aerial vehicles (UAV) and reconfigurable intelligent surfaces (RIS) can be employed for diversifying and enriching the security context of advanced wireless networks.  
Reference~\mbox{\cite{6Gcontextaware}} presents the security concerns in 5G and 6G with a visionary contextual framework that leverages AI to enhance the physical layer security. % in 6G. 
The authors of~\mbox{\cite{6GSecSurvey}} survey the key technologies of 6G and some of the security and privacy threats associated to them. % mentioned technologies and applications of 6G.
Reference~\mbox{\cite{6GSec}} discusses how 6G technologies can be used for enhancing the security of 6G networks and identify the corresponding research directions and challenges. %However, the current literature lacks a comprehensive study of security vulnerabilities and weaknesses across the chain of layers in 6G networks with feasible recommendations for empowering security aspects of the system.} 

This paper identifies the important 6G network threats %of  systems 
through AI and discusses how 6G technology can be leveraged to diversify security and design reliable and available networking solutions. Table~\ref{tab:survey} compares the key 6G technologies with those of 5G and the main security implications. % of the evolution towards 6G networks.
The focus of this paper is on AI attacks and mitigation.

%\hl{related works~\mbox{\cite{6Gcontextaware,6GSecSurvey,wang2020security}}}  \\
%\hl{If possible we should cite these two papers~\mbox{\cite{6GHorizons}}}

%Please include this \cite{tariq2020speculative} 
%\hl{Ref. 2 is related work}

The rest of the paper is organized as follows. Section II introduces major AI attack categories. Section III discusses the security threats associated with 6G technologies at different layers. 
Section IV elaborates on the opportunities of such technology to enhance network security.
Section V presents the ongoing standardization efforts towards secure cellular networks of the future and our recommendations, before deriving the conclusions.

\section{Artificial intelligence Attacks}
\label{sec:AIAttacks}
AI algorithms have found rich use cases and applications that will play as key enablers for the evolution of cellular networks towards 6G. % wireless communication network. 
The explosion of the number of users and connected devices with the integration of structured and unstructured network topologies %over such as space, aerial, and ground 
will yield extremely heterogeneous networks and services that will make the utilization of AI inevitable for 6G. % isfor network access, orchestration . The
AI driven solutions will be deployed in several domains and at different layers of  6G wireless networks. 
For example, modulation recognition, channel coding, channel estimation, authentication and identification, scheduling, and multiple access. 
In addition, deep learning (DL) based techniques have been adopted widely in wireless communications and networking due to its ability to solve parameter rich and computationally complex problems that do not have analytical %or infeasible
solutions. 

Regarding the security of wireless networks, the AI algorithms can act as a double-edged sword. On the defensive side, AI can be exploited for better understanding of attack patterns and enhance the discovery and recovery capabilities of wireless networks against security threats. On the offensive side, some of the attacks may take advantage of the learning process carried on by the AI to gain insights into user information, network operation, and even system management procedures for designing advanced attack strategies. 

The focus of this section is to cover the types of offensive AI attacks that the future 6G wireless networks will likely experience. 
Privacy in AI models can be violated via different attack strategies, such as extraction of training data, extraction of model information, and misprediction by employing an active or passive attack to manipulate the original data set. The attackers can develop %deploy
attack strategies as a function of the available information about the AI models used by one or several 6G network elements/layers. %at the attacker side .

\subsection{White vs. black box attack}
There are two main deployment cases for AI attacks. %, which are white box attack and black box attack. 
The white box attack targets the internal structure of the model and the intermediate process including the model parameters. On the other hand, the black box attack aims to make the AI model misbehave through the manipulations of input data due to the lack of observations about the model. The specific forms of attacks on the AI models that are discussed here are poisoning, reverse, and adversarial attacks. %The poisoning attack is an active attack that builds its strategy on targeting the training phase of the AI process via
%altering the raw training data and its probability distribution.
Reference~\cite{AIAttacks} provides further information related to the different types of AI attacks that can be applied to the wireless communications sector.

\subsection{Poisoning attack}
The poisoning attack is an active attack that builds its strategy on targeting the training phase of the AI process by
altering the raw training data and its probability distribution.
This is achieved %e poisoning attack becomes able to change the training dataset 
through the alteration of existing training data or the injection of mislabeled malicious samples into the original data set. The attacker may thus be able to modify the decision boundaries or reduce the accuracy or efficiency of the model. The result may include wrong outputs, misclasifications, or mispredictions %will be produced 
that can lead to poor 6G network %configuration
decisions. 

\subsection{Reverse attack}
The reverse attack is a passive attack that invades the testing phase of the AI model to extract private training data of the users or the model information. The reverse attack can be deployed in the white box style through utilizing the collected information about the AI model and reconstructing the original training data. In a black box style reverse attack the attacker steals the model features with the assistance of the application program interfaces (APIs). The reconstruction of row training data via the reverse attack can raise vital concerns regarding users' confidentiality and privacy violations. In addition, a reverse attack may work as a pre-processing mechanism to enable a harmful attack to the system by learning the network management policies and creating corresponding counter samples that violate the system precaution security algorithms. 
%The adversarial attack is a passive attack that fits in the testing cycle of the AI steps. 

\subsection{Adversarial attack}
The adversarial attack and the poisoning attack both have the same attack target, but the difference is what subprocess of the model that is being targeted. While the poisoning attack is being performed during the training phase, the adversarial attack takes place during the testing stage. The adversarial attack is capable of creating new adversarial samples that may increase the misclassifications; % recognize probabilities, but adversarial attack 
it does not alter the training data set. The adversarial attack can be implemented through the white box or black box strategy. 
\textcolor{black}{One type of common adversarial attack, the evasion attack, attempts to include samples that may appear legitimate to the system but are still harmful. This allows harmful samples to be contributed, while evading detection.}

\begin{table*}[h!]
\centering
\caption{6G vs. 5G RAN and the security implications. %\hl{I'm afraid we will need to introduce every acronym somewhere (text, caption, inside the table)}
}
\footnotesize
%\resizebox{0.85\textwidth}{!}{%
%\begin{tabular}{|c|c|c|c|}
{\begin{tabular}{|p{1.2cm}|p{5cm}|p{5cm}|p{5cm}|}
\hline
\textbf{Category} &  \textbf{5G} &   \textbf{6G} &  \textbf{Security Implications}
\\ \hline
\vspace{0.001 in}
PHY \& MAC &
\vspace{-0.05 in}
\begin{list}{\labelitemi}{\leftmargin=0.2em}
    \item {\textbf{Modulation schemes:} low to very high order. %BPSK, QPSK, 16QAM, 64QAM, 256QAM.
    }
    \item {\textbf{Multiple Access:} single-carrier frequency division multiple access (SC-FDMA) and orthogonal FDMA (OFDMA).}
    \item {\textbf{Waveform:} Adaptive lattice, windowing and filtering. %(OFDM, SC-FDE).
    }
    %\item {High MCS options.}
    \item {\textbf{Channel coding:} Turbo codes, low-density parity check (LDPC), and polar codes.}
    \item {\textbf{User-Cell association:} coordinated multipoint (CoMP) transmission/reception, Cloud RAN (C-RAN), small cell, and centralized massive multiple input, multiple output (mMIMO). }
    \vspace{-0.1 in}
\end{list}

&
\vspace{-0.05 in}
\begin{list}{\labelitemi}{\leftmargin=0.2em}
    \item {\textbf{Modulation schemes:} 5G traditional modulation options and Multi-domain modulation suhc as index modulation (IM) (shape, interval, position, etc.), space, time, frequency, etc.) and orbital angular momentum modulation.}
     \item {\textbf{Multiple Access:} free and full duplex communications. %(NOMA, SCMA)).
     }
    \item {\textbf{Waveform:} Adaptive windowing and filtering, orthogonal, non-orthogonal, single and multi-carrier, %(OFDM, GFDM, OTFS, FBMC, waveform-domain NOMA, 
    and multiple waveforms within a single radio frame.}
    %\item {Ultra adaptive and intelligent MCS.}
   \item {\textbf{Channel coding:} Advances polar codes, Enhanced FEC in the turbo codes, and serially or parallel concatenated schemes.}
   \item {\textbf{User-Cell association:} cell-free (CF) mMIMO.}
    \vspace{-0.1 in}
\end{list}

&
\vspace{0.001 in}
Protocol-aware jamming and spoofing attacks can lead to link adaptation failures or denial of service, among others. 
\newline Attacks to (RF) data, such as control and data channels used for AI model training, will compromise data driven channel estimation, link adaptation, channel access, and demodulation, among others. Using advanced channel access and waveforms for higher spectral efficiency, will make the wireless links more vulnerable to RF attacks.

\\ \hline
\vspace{0.001 in}
Frequency Bands &
\vspace{-0.05 in}
\begin{list}{\labelitemi}{\leftmargin=0.2em}
    \item {microWave}
    \item {mmWave}
    \vspace{-0.1 in}
\end{list}

&
\vspace{-0.05 in}
\begin{list}{\labelitemi}{\leftmargin=0.2em}
    \item {microWave }
    \item {mmWave }
    \item {THz }
    \item {Visible light }
    \vspace{-0.1 in}
\end{list}
&
\vspace{0.001 in}
The low penetration power of the THz signals will empower the jamming attacks capabilities weather through a physically blockage between the transmitter and receiver path or through creating environmental conditions,
such as gaseous particles or molecules.
\newline Optical beamforming jamming attack that can generate much higher illumination at the receiver than the legitimate signal by the transmitter.

\\ \hline
\vspace{0.001 in}
Network architecture &
\vspace{-0.05 in}
\begin{list}{\labelitemi}{\leftmargin=0.2em}
    \item {SDN/NFV}
    \item {Network slicing}
    \item {Service-based architecture (SBA)}
    \item {Cloud, fog, or edge computing architecture}
    \vspace{-0.1 in}
\end{list}

&
\vspace{-0.05 in}
\begin{list}{\labelitemi}{\leftmargin=0.2em}
    \item {Hybrid SDN/NFV architecture}
    \item {Adaptive and isolated network slicing }
    \item {Secured SBA architecture}
    \item {Cognitive service architecture (CSA)}
    \item {CF architecture}
    \item {Dynamic and integrated cloud/fog/edge computing architecture}
    \vspace{-0.1 in}
\end{list}

&
%\hl{Key points from Section III}
\vspace{0.001 in}
The increased decoupling of hardware and software will enable more non-proprietary options for implementing wireless networks and services. This will increase the attack surface and make unified and comprehensive initial checks and maintenance against known and unknown security threats more difficult.
\vspace{0.001 in}
\\ \hline
\vspace{0.001 in}

{Connec-tions} &%\cite{7} & %\begin{tabular}[c]{@{}c@{}}
%\begin{itemize}
%\leftskip .3in % see comments below
\vspace{-0.05 in}
\begin{list}{\labelitemi}{\leftmargin=0.2em}
%\leftskip .3in % see comments below
%\parindent -0.1in % see comments below
%\setlist[itemize]{leftmargin=*}
    %\item {\small UAV explorers}
    \item {Cellular}
    \item {Ad hoc}
    \item {Beam based}

    \vspace{-0.1 in}
%\end{itemize}
\end{list}

%  - Strong encryption mechanisms,
%  - Always encrypt user identifiers and location information
%  - Use of temporary identifiers that are rotated
&
% \begin{itemize}
%     \item Power Leakage
% \end{itemize}& % so that legitimate users cannot access available resources.%\end{tabular}
\vspace{-0.05 in}
\begin{list}{\labelitemi}{\leftmargin=0.1em}
\parindent -0.2in % see comments below
    \item {Cell free}
    \item {Dynamic ad hoc networking}
    \item {Multiple heterogeneous connections}
    \item {Satellites integrated network}
    \vspace{-0.1 in}
%\end{itemize}
\end{list}

&
\vspace{0.001 in}
Less structured RAN requires very reliable signaling and coordination among network access points, among mobile devices, and among both. 
\newline It needs trust in signals and data to access or provide access, which can be exploited by attackers.
\vspace{0.001 in}
\\ \hline

\vspace{0.001 in}
Services &

\vspace{-0.05 in}
\begin{list}{\labelitemi}{\leftmargin=0.2em}
    \item {Enhanced mobile broadband (eMBB)}
    \item {Ultra-reliable low latency communications (URLLC)}
    \item {Massive machine type communications (mMTC)}
    %\item {\small   Partial integrated of AI}

    \vspace{-0.1 in}
\end{list}
&
\vspace{-0.05 in}
\begin{list}{\labelitemi}{\leftmargin=0.2em}
    \item {Mobile broad bandwidth and low latency (MBBLL)}
    \item {Massive broad bandwidth machine type (mBBMT)}
    \item {Massive low latency machine type (mLLMT)}
     \item {Computation oriented communications (COC)}
      \item {Contextually agile eMBB communications (CAeC)}
       \item {Event defined uRLLC (EDuRLLC)}
     %\item {\small   Fully integrated of AI}
    \vspace{-0.1 in}
\end{list}

&
\vspace{0.001 in}
As services become more sophisticated and KPIs more demanding, they become more vulnerable. An attacker can cause system misalignment. For example, flooding the random access channel with requests can delay admission of latency-sensitive nodes. A slower response or occasional collision and packet drop in 5G may be a useless service in 6G.
\vspace{0.001 in}
\\ \hline

\vspace{0.001 in}
Security &

\vspace{-0.05 in}
\begin{list}{\labelitemi}{\leftmargin=0.2em}
    \item {Elliptic curve cryptography (ECC)}
    \item {SIM-dependent authentication }
    \item {Data-centric security}

    \vspace{-0.1 in}
\end{list}
&
\vspace{-0.05 in}
\begin{list}{\labelitemi}{\leftmargin=0.2em}
    \item {Post-quantum cryptography }
     \item {SIM-less authentication (embedded SIM (eSIM) and integrated SIM (iSIM))}
     \item {Data-driven security}
    \vspace{-0.1 in}
\end{list}

& 
\vspace{0.001 in}
Beyond protecting data from being exposed (privacy threat), the encrypted data needs to be protected against manipulation, tracking, inference attacks, so that authorized users can be efficiently authorized and trust be built and maintained.
%\vspace{0.001 in}

\\ \hline

%\vspace{0.001 in}
%Software vs. HW vs. openness &

%\vspace{-0.05 in}
%\begin{list}{\labelitemi}{\leftmargin=0.2em}
%    \item {\textbf{\small ...}}

%    \vspace{-0.1 in}
%\end{list}
%&
%\vspace{-0.05 in}
%\begin{list}{\labelitemi}{\leftmargin=0.2em}
%    \item {\textbf{\small ...}}
%    \vspace{-0.1 in}
%\end{list}

%\\ \hline

\vspace{0.001 in}
Intelligence &

\vspace{-0.05 in}
\begin{list}{\labelitemi}{\leftmargin=0.2em}
    \item {Partially integrated with a limited functions}

    \vspace{-0.1 in}
\end{list}
&
\vspace{-0.05 in}
\begin{list}{\labelitemi}{\leftmargin=0.2em}
    \item {Fully integrated at massive scale, across applications/services, and layers}
    \vspace{-0.1 in}
\end{list}
&
\vspace{0.0005 in}
The learning, self-adaptation and self-healing processes %of the 6G system 
need to be robust. If the data used for learning are compromised or manipulated, the system can be influenced to behave in a certain way that serves the purpose of the attacker.
%\vspace{0.001 in}

\\ \hline

\end{tabular}%
}

\label{tab:survey}
\end{table*}

\section{6G Enabling Technologies and Security Threats}
\label{sec:6GranSec}
In this section, we discuss the security threats in 6G considering top-down approach, from the network level to the physical (PHY) layer. Since AI will be an integral part of 6G, security a %revisit as
diverse range of attacks will be possible and will pose greater threat to the overall system. 

%Table \ref{tab:survey} describes the 

\subsection{ Network technology and attacks }
   
\label{subsec:NetworkTechnology}
%\subsubsection{Security threats to SDN/NFV in 6G}

Beyond 5G and 6G systems are witnessing architectural transformations in order to support service heterogeneity, coordination among multiple technologies and, rapid on-demand service deployment. %\cite{9247527}. 
Such transformations are enabled by the emerging open radio access network (O-RAN) framework which provides virtualization, intelligence, and flexibility while defining %. It is worth mentioning that the objective of O-RAN is to enhance the RAN performance through virtualization of network elements and 
open interfaces for network innovation. % that are complemented with intelligence. 
The distinguishing characteristic is the network intelligence that is %going to be 
the pillar of O-RAN and 6G wireless networks for rapid deployment, operation, and optimization. 
%the provisioning of optimization,  and rapid deployment.

\begin{figure}[t]
    \centering
    \includegraphics[width=0.48\textwidth]{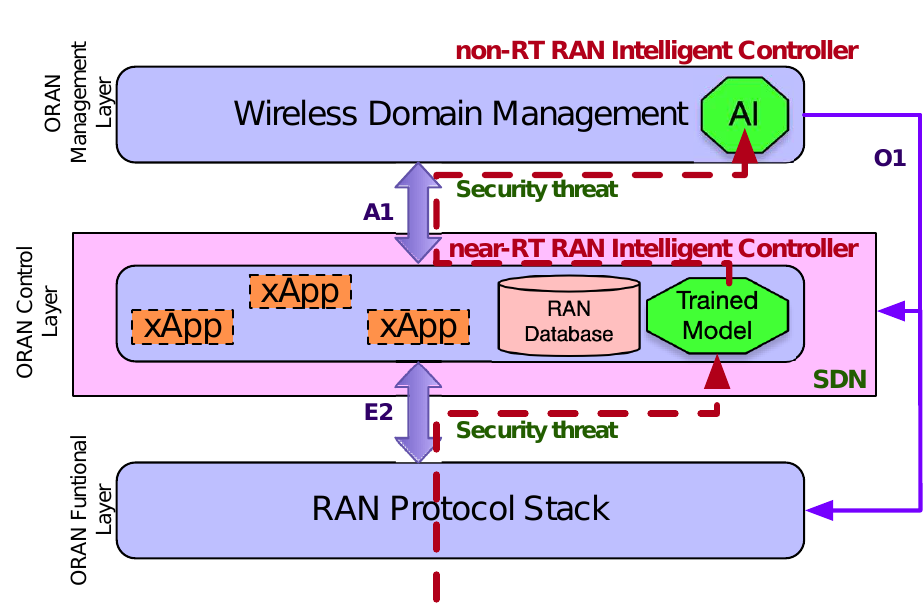}
    \caption{SDN-based O-RAN architecture with security threats at different levels.}
    \label{fig:SDN}
    %\vspace{-3mm}
\end{figure}

O-RAN makes use of software defined networking (SDN) and network function virtualization (NFV) technology enabling disaggregation of traditional RAN functions and connecting them using standardized interfaces. 
Fig.~\ref{fig:SDN} illustrates the SDN-based O-RAN architecture. %with %, as described in \cite{oRANalliance}, 
%is shown in . 

On the one hand, the disaggregation improves the security agility, adaptability and resiliency; on the other hand, it increases the security threat at the SDN level. An xApp is a microservice that is deployed by an equipment vendor, a software vendor, a network operator, or the open source community, and it requires robust security mechanism. It can be seen from Fig.~\ref{fig:SDN} that xApps monitor RAN operations by using the analytics from the non-real-time RAN  intelligent controller (non-RT RIC) using the A1 %open 
interface.

%The SDN is the key entity for providing the flexibility as it breaks the vertical integration (control and data planes) in traditional networks through centralized network control. In SDN, the control function is logically centralized at one or multiple control entities called SDN controller, and the data plane is simplified for network applications and services that the SDN is responsible for. Radio intelligent controllers(RICs), non real-time (non-RT) and near real-time (near-RT), are the intelligent O-RAN architecture modules (Fig~\ref{fig:SDN}). Both the modules work with AI training data to devise policies for deployment scenarios \cite{oRANallianceAI}.
%\textcolor{green}{
%AI in the non-RT RIC may be in charge of higher level functions such as capacity, frequency reuse, and management of the near-RT RIC. In the near-RT RIC, xApps will use AI for more finite functions in a modular manner. Some functions AI can be used for may include handovers, load balancing, and scheduling policies. 
%These xApps may also be deployed by third-parties, rather than directly by the operator. 
\textcolor{black}{Because of the versatility in creating xAPPs, the possibility of malicious deployment %of xApps has been increased
needs to be considered. Furthermore, AI at the non-RT RIC is still susceptible to standard attacks such as poisoning attacks, which can be used to deviate models from their optimal performance. Similarly, attackers may leverage AI for predicting the functions or outcomes from the AI algorithms at the RIC allowing attacks to be planned out based on the expected behavior. The open-source nature of O-RAN poses a double-edged sword where the software is more thoroughly examined and contributed to by a broader community, but a seasoned adversary can investigate the software for vulnerabilities and attempt to contribute improvements with built-in backdoors. Open interfaces also increase the surface that can be attacked. One such attack includes scanning the interfaces and using AI to identify vulnerable properties of the interface for attacking.}

%6G O-RAN evolves around the principle of intelligent RAN virtualization which inherits deployment-specific security challenges of SDN controllers. %as well as security problems raised by ML/AI algorithms.
%Besides the conventional security challenges in SDN as mentioned in \cite{ahmad2018overview}, %some of the 
%more advanced %possible
%security threats are: 
% \begin{itemize}
%   \item Lack of security and authentication on the north A1 interface between control and management layers, thus inviting spoofing attacks that can alter AI model description files. % sent to non-RT RIC from near-RT RIC. 
%   \item Inappropriate authorization and the absence of encryption of traffic on the south E2 interface, causing malicious access to applications as well as the ML training host. % at control layer. This is due to absence of encryption of traffic on south E2 interface. 
% \end{itemize}
AI is expected to improve several functionalities that SDNs will be responsible for. Among these are load balancing, resource utilization, fault detection, and security. %SDN controllers have the greatest level of access to the system making them a high value target to attackers. 
Given the span that SDN will be involved in 6G systems, they can become a high value target to attackers. 
Therefore, it is essential to acknowledge threats imposed both by AI attacks and by implementing AI into the system.

Implementing AI into SDN will improve performance in 6G networks. However, using AI presents some vulnerabilities that can be difficult to overcome. Existing attacks such as poisoning and evasion attacks can be difficult to detect, mitigate, and recover from. SDN AI will be in charge of optimizing load balancing and resource utilization. A typical poison attack may negatively influence the AI to decrease its effectiveness causing a drop in quality of service (QoS). Similarly, evasion attacks may be able to identify the vulnerabilities in a load balancing algorithm, allowing an attacker to take advantage of the system from the outside and decrease the QoS for outside users. The involvement of AI in fault detection and security poses related threats where a poisoning attack may cause the system to be unable to detect faults or security risks or even implement patches that further degrade security.

%% Vuk: already said
%%While AI has potential to improve the system itself, attackers will also attempt to leverage AI to degrade system performance. 
It is well known that AI can be used to help prevent attacks caused by software errors, day one vulnerabilities, and more. However, while AI can be used to effectively prevent these types of attacks, it can also be used maliciously to exploit these threats. Any interface that are accessible from outside the system may be vulnerable to an AI attack. Depending on the type of vulnerability, an adversary can deploy an attack such as through SDN applications, or a patterned distributed denial of service (DDoS) attack, or several other possibilities, including a series of attacks, to bring down a system. 
An SDN application, for example, may not seem malicious when deployed, but if it is aware of a hole in the system’s defense, it may pose a significant threat.
It is important to address the aforementioned security challenges of intelligent RANs, because malicious applications may otherwise be able to hijack the control flow and corrupt the data of the AI system, causing DoS attacks, among others.

\subsection{ Physical layer technology and attacks }
\label{subsec:PHYTechnology}
%Table \ref{tab:ORANuse} summarizes representative use cases of AI to configure and optimize lower layer network functions in the O-RAN framework.
This section focuses on attacks to two PHY layer functions that are the core of AI-enhanced advanced wireless communications.
\begin{figure}[t]
    \centering
    \includegraphics[width=0.48\textwidth]{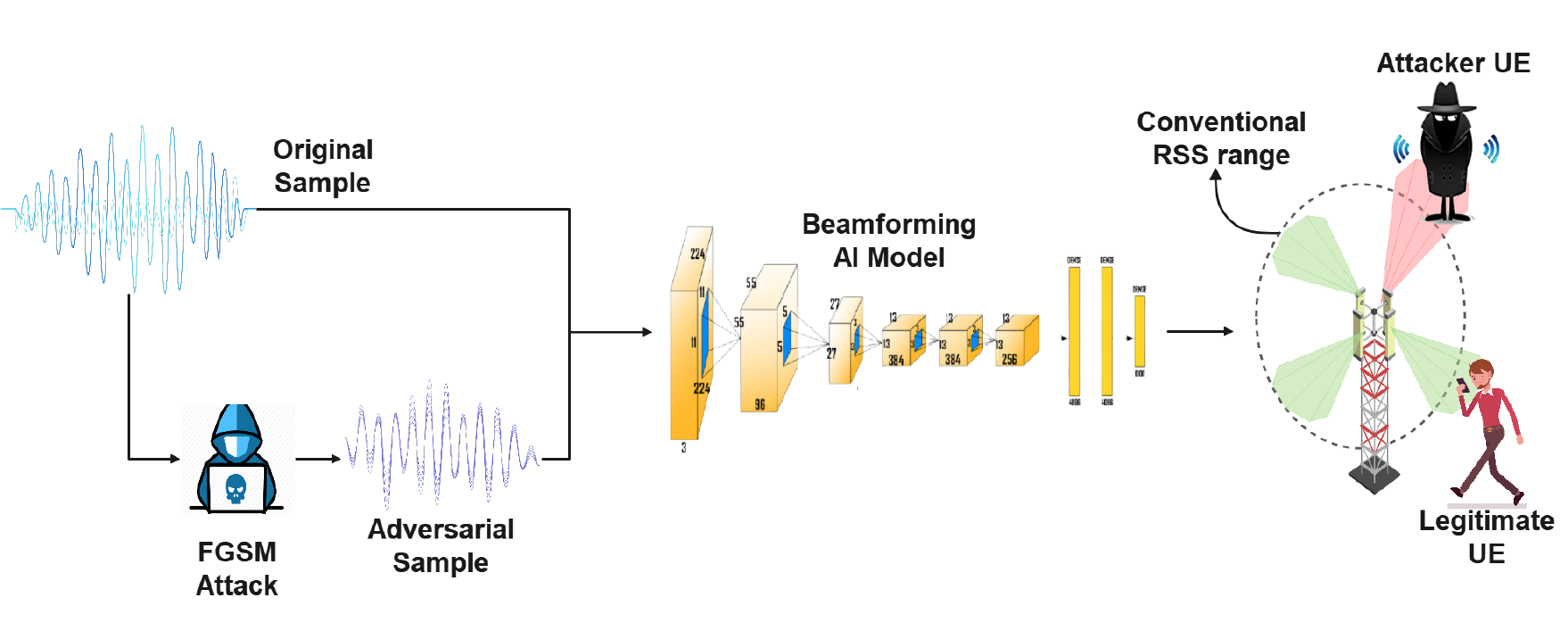}
    \caption{The FGSM attack on the beamforming prediction AI model.}
    \label{fig:FigureBeamAttack}
    %\vspace{-3mm}
\end{figure}
\begin{figure}[t]
    \centering
    \includegraphics[width=0.48\textwidth]{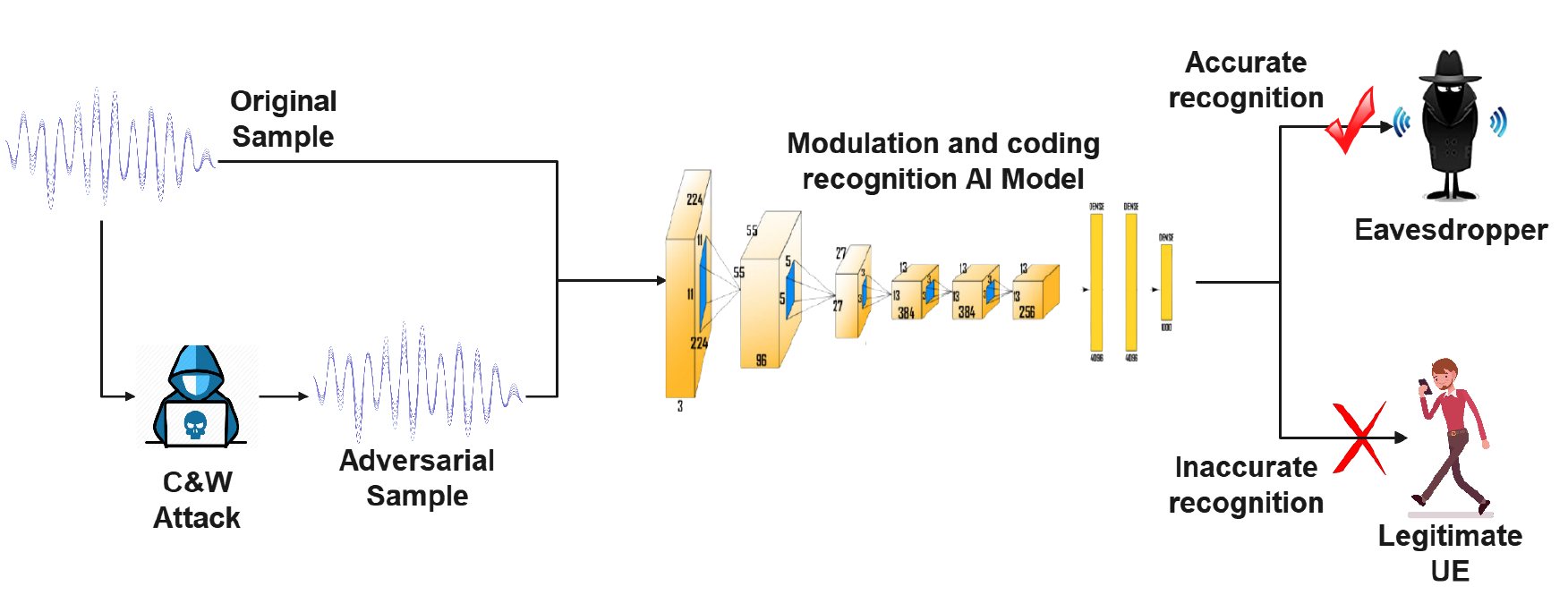}
    \caption{The C\&W attack on the AI model for modulation and coding scheme.}
    \label{fig:FigureMCSAttack}
    %\vspace{-3mm}
\end{figure}
% \begin{figure}[ht]
% \vspace{-5mm}
%   \subfloat[]{
% 	\begin{minipage}[c][0.45\width]{
% 	   0.5\textwidth}
% 	   \centering
% 	   %\vspace{5mm}
% 	   \includegraphics[width=0.97\textwidth]{figures/NextG/AI_AdversarialAttack_BeamForming.png}
% 	    \label{fig:FigureBeamAttack}
% 	    %\vspace{-35mm}
% 	\end{minipage}}
%  \hfill 	
%   \subfloat[]
%   {
%   %\vspace{-12mm}
%   %\hspace{2mm}
% 	\begin{minipage}[c][0.4\width]{
% 	   0.5\textwidth}
% 	   \centering
% 	   %\vspace{-10mm}
% 	   \includegraphics[width=0.97\textwidth]{figures/NextG/AI_AdversarialAttack_MCS.png}
% 	    \label{fig:FigureMCSAttack}
% 	    %\vspace{-35mm}
% 	\end{minipage}}
% %\vspace{-4mm}
% \caption{Illustration of the FGSM attack on the beamforming prediction AI model (a) and C\&W attack on the AI model for modulation and coding scheme (b).
% %\vspace{-3mm}
% }
% \end{figure}
\subsubsection{Beamforming Prediction}
The integration of AI into wireless communications can % 6G shows great potentials on 
enhance physical layer functions such as the beamforming design to provide the optimized beam directions for both the receiver and transmitter having multi-antenna arrays.The optimization of the beamforming vector is an essential element %that must be concerned in the future 
of 6G wireless networks t%o deploy different number of
for short and long-range communications techniques with rediscovered spectrum opportunities such as mmWave, Terahertz (THz), sub-6 GHz, and visible light communication (VLC). The AI techniques enrich the capability of providing exact beamforming vectors by feeding the RF information to different DL models for enormous number of base stations and users. 

However, accurate channel information is necessary for beamforming and %the security shortcomings of 
not being able to verify the channel information/RF data, especially for training an AI model, %s for enhancing %the prediction of 
%beamforming rise an alarm about the possible future security attacks where the attacker 
can be exploited to manipulate the beamforming prediction model through an adversarial or poising attack. 
The AI attacker may craft input streams that intend to fool the AI model to produce a failed or diverted beamforming vector in the favor of the attacker. The fast-gradient sign method (FGSM) %can be considered as one of the forthright with 
is such an %influential impact 
adversarial attack on the AI models that can be applied here~\cite{FGSMAIAttack}. The FGSM is capable of %circumventing 
manipulating the AI models for beamforming in 6G wireless networks by %creating adversarial samples 
by calculating the gradients of the loss function of the neural network. The adversarial samples %instances designed by the FGSM is
are crafted in such a way to maximize the loss value of the classifier hypothesis. % by adding the noise vector produced by the sign of the gradients to the original samples. 
The critical challenge that the FGSM faces is that the attacker must gain access to the AI model from the outside to produce the adversarial examples. Fig.~\ref{fig:FigureBeamAttack} illustrates how the FGSM can influence the performance of the beamforming prediction AI model to provide unintended beams with higher received signal strength (RSS) at the desired devices that are controlled by the attacker. 
\newline
\subsubsection{Modulation and coding scheme recognition}
Deep neural networks (DNNs) ae used extensively in the field of image recognition, speech analysis, and natural language processing. 
Its properties have been recognized for covering various applications in wireless communications, such as  channel decoding, modulation recognition, signal detection, and end-to-end communication system design. 
Adversarial attacks on DNNs can %discover its vulnerabilities and 
cause severe system performance degradation of the learning processes, moving classification decision boundaries that can lead to wrong decisions and degrade the accuracy of the AI model. 

The Carlini \& Wagner (C\&W) attack is another AI attack that has been developed to generate adversarial examples while taking into consideration that the crafted adversarial samples have the minimum perturbation size compared to the original samples~\cite{CWAIAttack}. 
C\&W is considered as one of the fewer AI attacks that is capable of generating less distortion of input samples as an adversarial sample for the same attack accuracy rate. 
Attackers may apply such techniques to ensure the correct recognition of the modulation and coding scheme (MCS) at the eavesdropper %side
while deceiving the DNN-based recognition models applied at the legitimate nodes to enforce false MCS recognition, which will effectively improve the eavesdropping rate over the user rate.
Fig.~\ref{fig:FigureMCSAttack} illustrates the procedure and goal of the C\&W approach to generate adversarial samples that are almost identical to the original samples with the ability to use it for fooling the trained DNN classifier employed at the legitimate receiver and prevent the accurate recognition of the modulation and coding scheme, for instance. 

Unlike the FGSM, the C\&W attack does not define the perturbation size as a default constraint, but as a variable that must be optimized for enriching the attack success rate. 
As a result of that, the C\&W attack has shown higher attack accuracy rate over the FGSM on the targeted DNNs with the possibility of performing a successful attack on the defensive distillation DNNs, where the train phase is done through the transferred knowledge of other DNN to reduce the computational complexity. On the other hand, the FGSM is computationally simple to develop because it depends on a single back-propagation computation compared to the costly optimization loop that the C\&W method performs for creating adversarial samples.

%\newline
Table \ref{tab:SecurityComparison} compares the traditional to AI attacks at the PHY and SDN/NFV layers. It shows that the AI attacks add complexity and severity to secure and reliable 6G network operation.

\begin{table*}[]
\centering
\caption{Comparison between security threats of conventional and AI-based attacks and their severity.}
\label{tab:SecurityComparison}
\footnotesize
\begin{tabular}{p{1.2cm}|p{5cm}|p{5cm}|}
\cline{2-3}
& \textbf{Physical Layer} & \textbf{SDN/NFV}       
\\ \hline
\multicolumn{1}{|c|}{}                                       
& \cellcolor[HTML]{FFFFFF} 
\begin{list}{\labelitemi}{\leftmargin=0.2em}
    \item {Malicious users mimicking signals from legitimate users and decreasing the secrecy rate %SNR
    through beamforming.}
    \item {By knowing the physical environment/channel between legitimate transmitter and receiver, physical features such as modulation and coding can be extracted by an eavesdropper.}
    \item {Core-level encryption techniques can easily block such users from %making catastrophic 
    affecting the network.}
    \item {\textbf{Severity}: Low}
\end{list}   
% & \cellcolor[HTML]{FCFF2F}{\color[HTML]{333333} }                      
& \cellcolor[HTML]{FFFFFF}{\color[HTML]{FE0000} }
\begin{list}{\labelitemi}{\leftmargin=0.2em}
    \item {Getting hold of the IP address of an SDN controller, attacks like DDoS, spoofing attacks, controller resetting, can be initiated. %by attackers.
    }
    \item {SDN/NFV is regarded as the brain of the network and any potential risk can cause irreversible damage to the network.}
    \item {\textbf{Severity}: Very High}
\end{list}
\\
\multicolumn{1}{|c|}{\multirow{-15}{*}{\textbf{Conventional Attacks}}} 
& \multirow{-3}{*}{\cellcolor[HTML]{FFFFFF}}
% & \multirow{-3}{*}{\cellcolor[HTML]{FCFF2F}{\color[HTML]{333333} }} 
& \multirow{-3}{*}{\cellcolor[HTML]{FFFFFF}{\color[HTML]{FE0000} }} 
\\ \hline
\multicolumn{1}{|c|}{}  
& {\cellcolor[HTML]{FFFFFF}}
\begin{list}{\labelitemi}{\leftmargin=0.2em}
    \item {Attack can create signal to manipulate AI model that produce beamforming vector in the favor of attacker.}
    \item {Attacker can disguise itself as a legitimate user %to the AI model 
    to get information related to physical layer, hence decreases the secrecy rate. %SNR.
    }
    \item {Because 6G will have a diverse range of services, physical layer attacks to the AI model can wreck havoc in the system.}
    \item {\textbf{Severity}: Very High}
\end{list}
% & \cellcolor[HTML]{FCFF2F}{\color[HTML]{333333} }                      
& \cellcolor[HTML]{FFFFFF}{\color[HTML]{FE0000} } 
\begin{list}{\labelitemi}{\leftmargin=0.2em}
    \item {AI-based xApps developed by an untrusted third party increases the risk to the AI trained model that can severely disrupt the network functions.}
    \item {\textbf{Severity}: Very High}
\end{list}
\\
\multicolumn{1}{|c|}{}  
& {\cellcolor[HTML]{FFFFFF}} 
% & \cellcolor[HTML]{FCFF2F}{\color[HTML]{333333} }           
& \cellcolor[HTML]{FFFFFF}{\color[HTML]{FE0000} }                   
\\
\multicolumn{1}{|c|}{\multirow{-15}{*}{\textbf{AI-based Attacks}}}     
& {\multirow{-3}{*}{\cellcolor[HTML]{FFFFFF}}} 
% & \multirow{-3}{*}{\cellcolor[HTML]{FCFF2F}{\color[HTML]{333333}}} 
& \multirow{-3}{*}{\cellcolor[HTML]{FFFFFF}{\color[HTML]{FE0000}}} 
\\ \hline
\end{tabular}
\end{table*}

\section{%Possible 
countermeasures to 6G-AI security challenges and research directions}
\label{sec:counterMeasures}
Appropriate models and simulators need to be developed to implement and quantitatively assess the security threats to AI-enabled 6G RANs and the effectiveness of countermeasures. While a preliminary assessment can be done in isolation, a true evaluation will require Beyond 5G network testbeds for experimental verification of both the feasibility of attacks and the proposed solutions. Candidates testbeds are the Platforms for Advanced Wireless Research which deploy software-defined cellular network testbeds that are combined with commercial network equipment. 
These testbed allow for implementing the O-RAN interfaces and a framework for security research in a controlled, yet production-like environment. In the reminder of this section we identify the technology for R\&D to address the previously identified attacks.  
\subsection{Security solutions for SDN/NFV}
\label{subsec:SDNsolutions}
For the O-RAN/SDN architecture shown in Fig.~\ref{fig:SDN}, a robust authentication mechanism is needed to protect the application as well as the AI models at the control and management layers.
%A proactive approach is to devise an authentication mechanism that restrict malicious applications.
Quantum key distribution (QKD) and %key
delivery processes can be centrally managed by the SDN controller which can demand keys from all %malicious
applications to identify the legitimate and restrict the malicious ones. 
Similarly, VNFs must have trust mechanisms to ensure proper operations. % and outputs them. 
We recommend continuous AI methods as embedded security functions in containerized VNFs for attack prevention, detection, and mitigation, such as through traffic monitoring. 
Isolation of containers also helps prevent outside attacks and improved resource management.
%A reactive approach is to analyze the data flow through the application programming interfaces (APIs) and employ AI models %can be incorporated in order 
%to capture threats coming from malicious applications.

Since third party Cloud platforms will be shared among different telecommunications operators, insider attacks need to be considered. O-RAN provides the facility where the switches store the data plane flow behavior and flow statistics, which the SDN controller can easily fetch without increased overhead. 
This would enable %As a result, 
network-wide consistent security policies and monitoring of the input data flow and the AI model output for dynamic protection of the AI controllers. Additionally, AI techniques should be considered for other wireless network security mechanisms, including authentication, authorization, encryption, and validation. %should be adopted in order 
\subsection{Security solutions for the physical layer}
Most of the physical layer security attacks mentioned in the previous section rely on the use of carefully crafted adversarial examples to mislead the AI models and disturb the wireless communications performance. % of the physical layer techniques. 
Therefore, one of the most common defense mechanisms that shall be deployed for mitigating the effect of adversarial attacks and ensuring robustness of the AI models is adversarial training. 
The adversarial training is the process of reproducing adversarial samples by following the same strategies of adversarial attacks and retrain the AI model with the friendly adversarial samples and its labels. 
This process aims to reshape the input streams of the AI model with unexpected samples to create sophisticated structures of input samples and prevent the adversary from defining those. 
Randomized smoothing can also be applied alongside with the adversarial training to enhance the robustness of the classifiers specifically in testing phase. 

The critical challenge of such adversarial training %mitigation technique 
is the possibility of using a different adversary strategy than the future attacker. % side different than the strategy applied by the adversarial training. 
In addition, %tin the training phase. T
the use of friendly adversaries during the training phase will degrade the performance of the AI model. 
These challenges can be defeated through applying various adversarial attack strategy in the training phase and the use of the adaptive regularizer that can provide a high classification accuracy. % under the existence of friendly adversarial examples. 

\subsection{Blockchain-based security solutions}
\label{subsec:blockchainsol}
 The nature of 6G networks where a massive number of heterogeneous devices and connections will coexist and share %the same
 spectrum and other radio resources emphasizes on the need for a %good and secure
 distributed security approach. The intrinsic characteristics and security mechanisms of the blockchain technology %, as discussed in \ref{subsub:Blockchain}, 
 enable decentralized %security, %solutions and %it a 
  resource and spectrum management. %approach for the 6G networks. 
%% One %of the 
%% area where %to integrate
%% blockchain %as the underlying infrastructure 
%% can  provide a secure, salable and efficient solution is spectrum sharing. Another application  where blockchain can provide a secure %challenges
%% solution is UAV communications. 
In other words, with everything being connected in 6G, %, including UAVs, 
blockchain can provide the necessary decentralized and trustworthy communications infrastructure. %for the UAV network. 
 For network virtualization and slicing, the smart contracts feature %in the blockchain 
 can be used for slice trading and management of the service level agreement (SLA) where the integrated blockchain can %be integrated for 
 record how each resource has been used and how each service provider has performed with respect to the SLA. Such %The
 integration %of blockchain in network slicing 
 can provide transaction and coordination cost savings, in addition to security and operational efficiency for each slice. Also, the %blockchain
 distributed ledger can be used %for AI applications 
 as a storage scheme to ensure integrity of the training data %and to secure it from %any malicious 
 against tampering. 
 
 The training itself needs to be done in a secure, yet production-like environment. %, which a production environment may not be.
 Research is needed to determine where/when to collect training data and where/when not, how to ensure enough redundancy, verification, and sanity checks to detect and discard potentially corrupted data. % the data in this regard, for example, when  and how to use redundancy to sanity check the data is needed.

\section{Standardization Efforts and Recommendations}% for 6G Security}
\label{sec:standardizationSection}

%\textcolor{red}{\hl{split in different paragraphs and elaborated/make a recommendation in each}}

%\hl{6G standardization, 3GPP attacks, reports about cellular network attacks, ...}

\begin{figure}[t]
    \centering
    \includegraphics[width=0.48\textwidth]{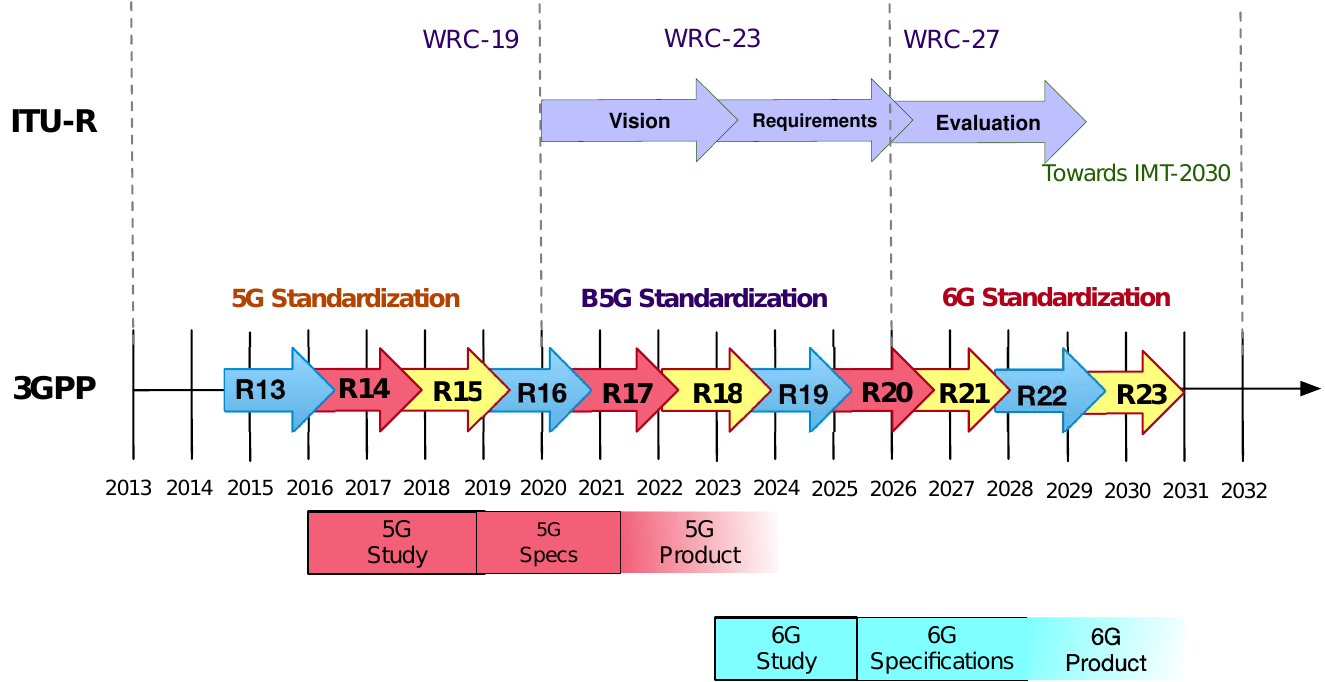}
    \caption{6G standardization timeline.}
    \label{fig:6Gtimeline}
    %\vspace{-3mm}
\end{figure}

The Third Generation Partnership Project (3GPP) Release 17 (Rel-17) features are in the standardization phase and most likely will be standardized by the end of 2021. ZTE and Samsung expect that 6G will be standardized by the 2029 %with the features of 
in 3GPP Rel-21 and will be commercialized by 2030. This timeline is depicted in Fig. \ref{fig:6Gtimeline}. 
\textcolor{black}{The International Telecommunication Union - Radiocommunication Sector (ITU-R) organizes the World Radiocommunication Conference (WRC) every 4 years to discuss about spectrum related issues. It is expected that WRC-23, in 2023, will discuss 6G related spectrum allocation that will pave the way for 6G spectrum allocations in WRC-27. %in 2027. 
Spectrum allocation will formally be decided by considering different aspects of the network, where reliability and security are of particular importance.} 
% Certain features that have already been proposed for 5G but due to non-availability of advanced electronics, those features will be included in 6G, e.g., advanced waveforms, modulation and coding, encryption techniques, etc. 
The security concerns are likely to be revised for 6G with the appearance of quantum computers which will require to define %  ``\textit{Intelligent}'' 6G network demands 
more tightened end-to-end security constraints at all layer of the network. 

Security of the 6G RAN will be the focus of research in the years to come. Standardization bodies, including the 3GPP, the ITU-R, the European Telecommunications Standards Institute (ETSI), and ISO/IEC Joint Technical Committee (JTC-1), have already started working on standardizing AI for 5G and beyond wireless systems. For instance, a logical architecture for high-level requirements is defined by ITU-T Rec. Y.3172 \cite{recommendation3172}. One of the key aspects of such an architecture is the technology independence which means it can be easily integrated into current and future technologies without interoperability issues.  

However, these standardization bodies are skeptical about standardizing the AI models and algorithms. Because 6G encompasses a wide range of applications with massive connectivity (M2M/IoT), it is required to define signaling support for training and execution of AI-based security mechanisms, considering the constraints imposed by light-wight devices which have limited storage and power. 

In order to enable %a complete
zero-trust security mechanisms, the standardization bodies are required to insert PHY layer security into the standards. Since the 3GPP is heavily relying on the protocols/algorithms of the Internet Engineering Task Force (IETF), defining PHY layer security as a lightweight security mechanism for low-power devices will open new horizons for researchers, developers and vendors. The next generation mobile networks (NGMN) will therefore need to devise SDN and NFV security protocols that will be proactive as well as reactive to security threats, unlike in 5G systems. % where protocols are expected reactive. 
For instance, proactive authentication mechanisms for encrypted interfaces and secured containers for VNFs are desired as 6G networks will dependent on softwarization. A reactive approach is to consider AI-based security mechanisms against security threats. The related study items for AI and novel encryption techniques have already been initiated and additional research %further study 
on these technologies will soon be in high demand. 

It is important to mention that end-to-end cryptography in the network should be considered in the context of post-quantum cryptography.
Both ETSI \cite{ETSI-WKD} and ISO \cite{JTC1-WKD} have started exploring QKD as post-quantum cryptography mechanisms. In particular, QKD interoperability among devices from different vendors is an issue that is needs to be address in the standards.  The national institute of standardization (NIST) is currently looking into post-quantum cryptography algorithms in order to make public key encryption schemes more resilient in the future.
Pre-standardization research %phase 
should consider the aforementioned issues related to 6G RAN security targeting zero-trust to minimize the potential losses of cyber attacks which will increase with the dependence on wireless connectivity. 

%\section{6G Enabling Technologies, Applications, and their Security Requirement and Threats}
%\label{sec:case}
%\input{./include/case.tex}

% \section{6G Enabling Technologies}
% \label{sec:Enabler}
% \input{./include/enabler.tex}

% \section{Proposed Strategies}
% \label{sec:contribution2}
% \input{./include/contribution2.tex}

%\section{Research Directions}
%\label{sec:case}
%\input{./include/directions_v2.tex}

\section{Conclusions}
\label{sec:conclusions}
This paper focuses on the security of 6G RANs which will rely on intelligence. We consider the O-RAN framework as the basis for AI-enabled 6G and elaborate on the critical security threats and potential solutions at the network and physical layers. The traditional cellular network attacks that exploit the protocol will shift to attacks to the AI models and the corresponding training data. While 6G networks are not expected until 2030, pre-standardization efforts, also related to security, have already started. 
Security research is needed to define and revise intelligent cellular communications technology, architectures and protocols, before, during and after initial standardization and deployment as operational and security contexts change.

\section*{Acknowledgement}
% 52 words included in overleaf count
This work was supported in part by the National Science Foundation, under grant number CNS-2016724.

\balance

\bibliographystyle{IEEEtran}
\bibliography{imsi_bib,vuk,UAVs,New_Ideas,aly}
% 168 words included in overleaf count
\section*{Biographies}
\footnotesize
\vspace{0.2cm}
\noindent
\textbf{Talha Faizur Rahman} (tfr42@msstate.edu) is a postdoctoral researcher in the Department of Electrical and Computer Engineering at Mississippi State University, Starkville, MS, USA. His research interests are on wireless communications, signal processing, and the Internet of Things.

\vspace{0.2cm}
\noindent
\textbf{Aly Sabri Abdalla} (asa298@msstate.edu)
is a PhD candidate in the Department of Electrical and Computer Engineering at Mississippi State University, Starkville, MS, USA. His research interests are on scheduling, congestion control and wireless security for vehicular ad-hoc and UAV networks.

\vspace{0.2cm}
\noindent
\textbf{Keith Powell} (kp1747@msstate.edu) is pursuing a PhD degree in the Department of Electrical and Computer Engineering at Mississippi State University, Starkville, MS, USA. His research interests include software radio platforms, UAV communications, and embedded systems.

\vspace{0.2cm}
\noindent
\textbf{Walaa AlQwider} (wq27@msstate.edu) is pursuing a PhD degree in the Department of Electrical and Computer Engineering at Mississippi State University, Starkville, MS, USA. Her research interests include spectrum sharing between active and passive systems and AI-enhanced communications.

\vspace{0.2cm}
\noindent
\textbf{Vuk Marojevic} (vuk.marojevic@msstate.edu) is an associate professor in electrical and computer engineering at Mississippi State University, Starkville, MS, USA. His research interests include resource management, vehicle-to-everything communications and wireless security with application to cellular communications, mission-critical networks, and unmanned aircraft systems.

\end{document}